# Chirality-driven ferroelectricity in LiCuVO$_4$

Alexander Ruff[1], Peter Lunkenheimer[1], Hans-Albrecht Krug von Nidda[1], Sebastian Widmann[1], Andrey Prokofiev[2], Leonid Svistov[3], Alois Loidl[1], Stephan Krohns[1,*]

[1] Experimental Physics V, Center for Electronic Correlations and Magnetism,
University of Augsburg, 86135 Augsburg, Germany
[2] Institute of Solid State Physics, Vienna University of Technology, A-1040 Wien, Austria
[3] P.L. Kapitza Institute for Physical Problems, RAS, Moscow 119334, Russia

**Chirality or the handedness of objects is of prime importance in life science, biology, chemistry and physics. It is also a major symmetry ingredient in frustrated magnets revealing spin-spiral ground states. Vector chiral phases, with the twist (either clock- or counter clock-wise) between neighbouring spins being ordered, but with disorder with respect to the angles between adjacent spins, have been predicted almost five decades ago. Experimental proofs, however, are rare and controversial. Here, we provide experimental evidence for such a phase in LiCuVO$_4$, a one-dimensional quantum magnet with competing ferromagnetic and antiferromagnetic interactions. The vector chiral state is identified via a finite ferroelectric polarization arising at temperatures well above the multiferroic phase exhibiting long-range three-dimensional spin-spiral and polar order. On increasing temperatures, spin order becomes suppressed at $T_N$, while chiral long-range order still exist, leaving a temperature window with chirality-driven ferroelectricity in the presence of an external magnetic field.**

In recent years, magnetic materials exhibiting spin spirals gained considerable attention due to their helical or cycloidal spin configurations[1]. There the chiral spin twist can induce spin-driven ferroelectricity[2] either via a spin-current mechanism[3] or via an inverse Dzyaloshinskii-Moriya interaction[4,5]. This revitalized the large field of multiferroics[6], being of great fundamental and technological importance[7,8]. Vector chirality, defined by the vector product of spins at adjacent lattice sites, $S_i \times S_j$, is one of the key concepts in quantum magnetism[9,10,11] and in one-dimensional (1D) spin systems. Chirality is also an important symmetry element of spin spirals. However, either finite temperatures or quantum fluctuations can completely suppress the long-range spin order of a spiral, while leaving the chiral twist less affected, leading to a vector chiral (VC) phase. For 1D XY helimagnets, forty years ago Villain[12] predicted such a phase appearing between the paramagnetic (PM) high-temperature state and the conventional long-range-ordered 3D helical spin solid.

The occurrence of a VC phase results from an exponential divergence of the chirality-chirality correlation with decreasing temperature (caused by the Ising character of the chirality order parameter), while the spin-spin correlation diverges via a power law. Based on similar arguments, Onoda and Nagaosa[13] proposed to search for chiral spin pairing just above the helical magnetic ordering temperature in 1D multiferroic quantum magnets. The ($H,T$) phase diagrams of $S = ½$ magnetic chains with competing interactions are complex. A variety of phases, including spin-density wave, vector chiral, nematic and other multipolar phases has been theoretically elucidated[14,15,16,17,18,19]. However, experimental explorations of these phases are rare. A possible validation of Villain's conjecture of a VC phase was reported, based on the two-step magnetic ordering observed in the quasi 1D helimagnet Gd(hfac)$_3$NITEt[20]. In light of the mentioned ferroelectric (FE) polarization induced by chiral spin twists[3,4,5], the most promising route seems to search for VC phases by identifying FE polarization at temperatures slightly above well-known 3D-ordered spin-driven multiferroic phases. From a theoretical point of view, quantum spin-chain systems with ferromagnetic (FM) nearest neighbour (nn) and antiferromagnetic (AFM) next-nearest neighbour (nnn) exchange, for which the inverse spinel compound LiCuVO$_4$ is a prime example, seem perfectly appropriate to detect the VC phase[13].

LiCuVO$_4$ offers a rich variety of complex magnetic phases, like the possible appearance of a bond-nematic phase at low temperatures and high magnetic fields[21,22,23,24,25]. In this system, the Cu$^{2+}$ ions carry a spin $S = ½$ and form a quasi-1D Heisenberg chain with competing nn FM exchange, $J_1/k_B \approx$ -19 K and nnn AFM exchange $J_2/k_B \approx$ 65 K[26]. However, strongly varying exchange constants are reported[27,28]. Due to weak inter-chain coupling, this frustrated 1D spin system undergoes 3D helical order at $T_N \approx$ 2.5 K[29,30], developing spin-driven ferroelectricity and, thus, multiferroicity[31,32]. Here, we report the occurrence of finite FE polarization in the PM phase of LiCuVO$_4$, just above the onset of 3D spin order. This FE phase, detected at moderate external magnetic fields, extends to temperatures up to five times $T_N$. Overall, our findings strongly point to the existence of a VC phase in the $S = ½$ quantum magnet LiCuVO$_4$, in accord with theoretical predictions[13].

Figure 1 shows results of heat-capacity and magnetic experiments on LiCuVO$_4$ at low temperatures and moderate external magnetic fields. The inset of Fig. 1a displays the temperature dependence of the magnetic susceptibility. The broad maximum is a fingerprint[33] of a $S = ½$ linear-chain compound with an average exchange of $J/k_B \approx$ -22 K[34]. The further increase towards low temperatures signals that the material approaches the magnetic phase transition at 2.5 K. The heat capacity, plotted as $C/T$ (Fig. 1a) exhibits a hump-like shape close to 12 K and reveals a characteristic λ-type anomaly when crossing the AFM and FE phase boundary at 2.5 K. For increasing magnetic field, the hump slightly shifts to lower temperatures while the AFM anomaly is almost temperature independent. The low-temperature phase boundary of the AFM and FE phase agrees well with the detailed ($H,T$) phase diagram published in ref. 32. However, there are no reports on the hump-like anomaly close to 12 K, weakly depending on the external field (see, e.g., ref. 34). Figure 1b shows the field derivative of the magnetization up to 5 T for a series of temperatures between 3 and 15 K. Remarkably, we find a well-established metamagnetic transition close to 1.5 T, which is best developed between 3 and 8 K. We ascribe this feature partly to the crystal anisotropy. By angular-dependent electron spin-resonance experiments, the anisotropic symmetric magnetic-exchange interaction of LiCuVO$_4$ was determined to be of the order of 2 K, corresponding to magnetic fields of about 2 T[30]. Moreover, a weaker but still significant anomaly appears close to 3.5 T. Beyond 8 K, the anomalies broaden and smear out; at 15 K they finally disappear. Interestingly, all the anomalies in heat capacity and magnetization appear within the PM and paraelectric phase, well above the onset of AFM and FE order, but still below the susceptibility cusp at ~ 25 K, characterizing the mean average magnetic exchange of a single spin chain. These anomalies, observed in magnetic and thermodynamic quantities, provide a first hint at the occurrence of a VC phase.

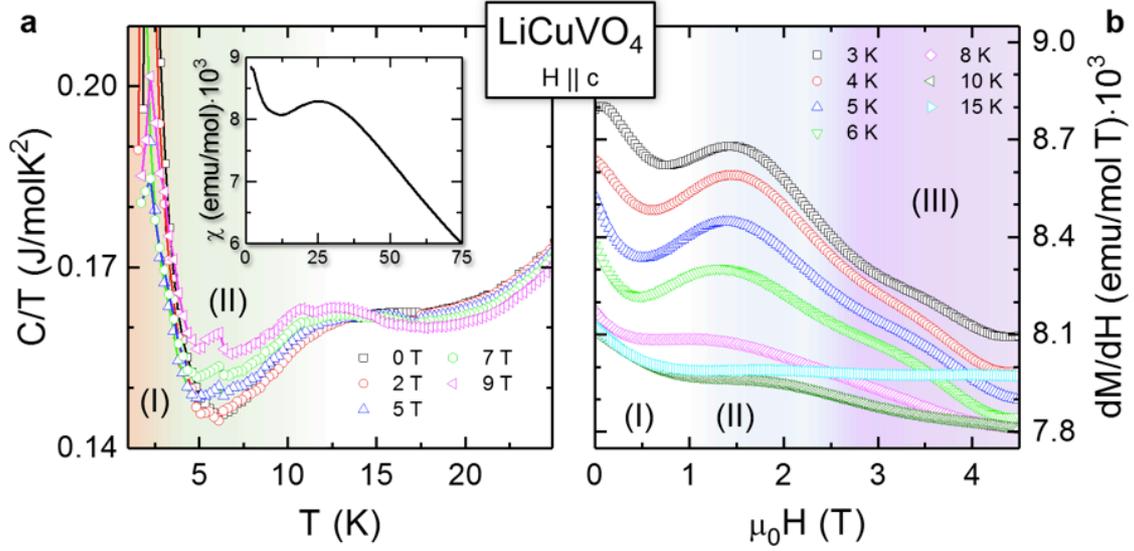

**Fig. 1 | Heat capacity and magnetization of LiCuVO$_4$. a**, Temperature dependence of the specific heat, plotted as $C/T$ for a series of external magnetic fields along the crystallographic **c** direction between 0 and 9 T. The inset shows the temperature dependence of the magnetic susceptibility as measured in an external magnetic field $\mu_0 H = 0.1$ T. (I) denotes the 3D magnetically ordered phase and (II) the regime of the field-induced VC phase, respectively. **b**, Derivative of the magnetization with respect to the magnetic field for fields $H\|\mathbf{c}$ up to 5 T taken for a series of temperatures between 3 and 15 K. For $3\,\text{K} \leq T \leq 15\,\text{K}$, (I) denotes the regime, where no magnetic or polar order is detectable, (II) the metamagnetic region without considerable macroscopic polarization and (III) the VC phase with distinct electric polarization.

As discussed above, experimentally the most convincing proof of a VC phase is to document the appearance of spin-driven FE polarization in the absence of long-range magnetic order. Figure 2 documents the polarization **P** in LiCuVO$_4$ as function of temperature (a) and magnetic field (b) measured along the crystallographic **a** direction in magnetic fields applied along **c**. The strong increase of the polarization below $T_N \approx 2.5$ K (Fig. 2a) signals the onset of spin-driven ferroelectricity (c.f. saturation polarisation $P_s \approx 30\ \mu\text{C/m}^2$ shown in Figure S2, Supplementary Information). At this temperature, a 3D long-range ordered cycloidal spin spiral evolves, propagating along **b** with the basal plane confined to the **ab** plane[26,31,32]. According to the symmetry rules of a spin-current mechanism[3] or an inverse Dzyaloshinskii-Moryia interaction[4,5], this spin configuration induces ferroelectricity with polarization along **a**[31,32,35]. However, Fig. 2a provides experimental evidence for a smaller ($\approx 0.1\ P_s$) but still significant ferroelectric polarization well above $T_N$, a possible hallmark of a cycloidal VC phase existing beyond the 3D ordered spiral phase[13]. In the VC phase, the twist is ordered, but the twisting angles between neighbouring spins are disordered, naturally explaining the much smaller polarization compared to the 3D ordered spin spiral with a coherent order of twist angles. This polarization evolves below 11 K and increases with increasing external magnetic field. However, in low magnetic fields ferroelectric polarization is absent within experimental uncertainty. To study this field effect in more detail, for temperatures between 3 and 13 K the field-dependent polarization is presented in Fig. 2b. It evidences the onset of marginal polarisation at 1.5 T and a significant increase for external fields $\mu_0 H > 3$ T. The inset in Fig. 2b shows the magnetic-field dependence of the dielectric constant $\varepsilon'$ at 4 K, revealing a small but well-defined peak at about 3.5 - 4 T, well above $T_N$. It represents a characteristic feature of improper ferroelectricity, in the present case induced by a metamagnetic transition (see Fig. 1b) and signals the transition into the long-range ordered VC phase. The question remains why no ferroelectric polarization exists at low magnetic fields. Most probably, on cooling well below the susceptibility cusp where the spin spirals are already well-developed, clock- and counter-clockwise spin chiralities are statistically distributed. This is certainly true for neighbouring spin chains, which are only weakly coupled, but could also be valid for spin fragments within one chain. We assign this phase with fluctuating spin chiralities as vector-chiral spin liquid (VCL). Obviously, an external magnetic field perpendicular to the basal plane of the spin rotation is needed to switch all chiralities in one direction. This is established at fields close to 4 T indicated by the strong increase of ferroelectric polarization and by the peak-like anomaly in the dielectric constant (see Fig. 2b).

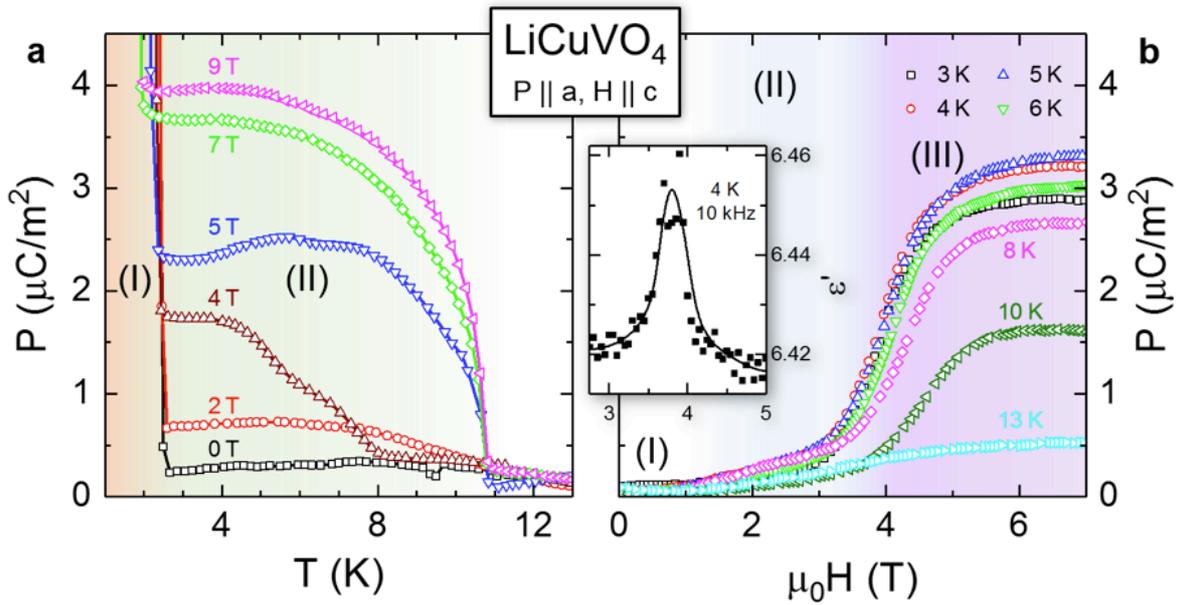

**Fig. 2 | Polarization of LiCuVO$_4$ in the region of the VC phase. a**, Temperature dependence of the electric polarization along the crystallographic **a** direction, $P\|\mathbf{a}$, for a series of magnetic fields $\mu_0 H$, between 0 and 9 T with $H\|\mathbf{c}$. (I) denotes the 3D magnetically ordered phase and (II) the regime of the field-induced VC phase, respectively. **b**, Field dependence of the polarization $P\|\mathbf{a}$ for a series of temperatures between 3 and 13 K. The inset shows the magnetic-field dependence of the dielectric constant measured at 4 K, well above $T_N$. (I) denotes the regime, where neither long-range magnetic nor polar order is detectable (VCL), (II) a transition region indicated by a metamagnetic anomaly with marginal polarization only and (III) the PM VC phase with distinct FE polarization.

These experiments – together with results obtained earlier[32] – allow for the construction of a detailed $(H,T)$ phase diagram of LiCuVO$_4$ including the 3D spin-spiral phases and the advocated VC phase extending to higher temperatures well into the PM regime (Fig. 3). Below 2.5 K, an AFM regime with long-range magnetic order and concomitant spin-driven ferroelectricity shows up. As discussed in detail in ref. 32, in the long-range magnetically ordered phase and at zero external magnetic field, the spiral axis **e** is oriented parallel to the crystallographic **c** direction and the spiral propagates along **b**, resulting in a polarization along **a**. Increasing magnetic fields along **c** leave the spin-spiral axis and, hence, the polarization unchanged. At 8 T the system undergoes a transition into a modulated collinear structure with zero polarization[32]. In the present work, we investigated the temperature and field regime up to 15 K and 9 T, identifying an extended region with finite ferroelectric polarization, above 3 T and up to temperatures of 11 K (Fig. 2). From these experiments, we conclude that the low-field regime is characterized by zero polarization, while for higher fields $\mu_0 H > 3$ T a significant polarization evolves. The natural explanation for the occurrence of this polarization is the existence of a PM VC phase in LiCuVO$_4$ developing at temperatures well above the onset of long-range magnetic order. This VC phase is characterized by long-range order of the spin chirality, however with statistical distribution of twist angles. When entering the low-temperature spin-spiral phase coherent order of the twist angles is established, providing long-range ordered spin correlations. Zero polarization in low fields probably results from fluctuating spin chiralities leading to a VCL phase. Between 1.5 T and 3 T we find an intermediate phase with marginal ferroelectric polarization. It could result from a distinct magnetic anisotropy[30]. But also could indicate that the phase diagram in LiCuVO$_4$ is even more complex with further chiral, multipolar or nematic-type phases as have been theoretically predicted[18,19]. Even domain-like objects of spin spirals with different rotation sense may play a role, leading to an extended temperature range for which vector chirality is stabilized[36,37].

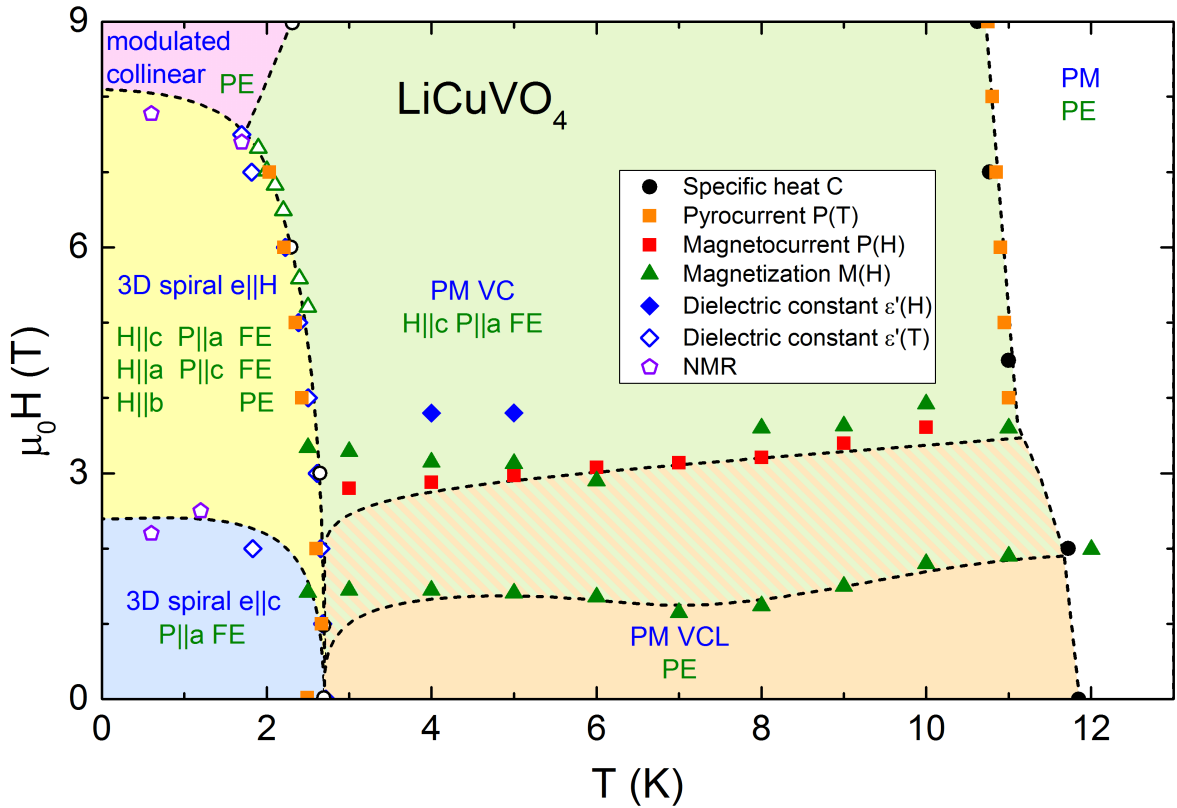

**Fig. 3 | ($H,T$) phase diagram of LiCuVO$_4$.** Results from the present work (closed symbols) and from ref. 32 (open symbols) were used. Anomalies observed by different measuring techniques are characterized by different symbols (see figure legend). Magnetic (blue lettering) and electric phases (green lettering) are indicated in the different regimes. The VC phase is indicated by the light-green area. This phase evolves at 3 T and extends at least up to 9 T. Its upper limit could not be identified in the present work. At low external fields up to 1.5 T a VCL phase with fluctuating chiralities is established. The intermediate phase (dashed area) could not be uniquely identified.

In summary, by measuring the polarization and dielectric constant in LiCuVO$_4$ as function of temperature and magnetic field, we provide experimental evidence of FE polarization in its PM phase, well above the well-known multiferroic phases. The existence of a VC phase with long-range ordered chiralities as a precursor of spin order is the most plausible explanation. Hence, our results provide the long-sought experimental hint of the theoretically predicted existence of a VC phase, in close analogy to cholesteric liquid crystals[12]. Moreover, we confirm the expectation that chiral spin pairing without long-range spin order should occur prior to 3D spin-spiral order[13] and demonstrate that the spin twist in a VC phase induces ferroelectricity, just as in conventional spin-spiral magnets, however, with much lower polarisation. It would be highly interesting to perform neutron-scattering experiments on LiCuVO$_4$, specifically focusing on the occurrence of this VC state.

**Methods**

LiCuVO$_4$ single crystals have been prepared as described in ref. 38. They crystallize in the space group *Imma* and undergo a transition into an AFM phase at ~ 2.5 K. The crystals have typical sizes of 3 × 3 × 1 mm$^3$ and were oriented by Laue diffraction techniques. The single crystals used in the course of this work were characterized by magnetic susceptibility and magnetic resonance techniques[30,39], as well as by dielectric techniques[32,35]. For detailed magnetization and heat-capacity experiments, we utilized a Quantum Design SQUID magnetometer with external magnetic fields up to 5 T and a Quantum Design Physical Property Measurement System (PPMS) using magnetic fields up to 9 T. The dielectric measurements were performed with electric fields applied along the crystallographic directions, with silver-paint contacts either in sandwich geometry or in a cap-like fashion, covering the opposite ends of the sample. The complex dielectric constants were measured for frequencies between 100 Hz and 10 kHz employing an Andeen-Hagerling AH2700A high-precision capacitance bridge. For measurements between 1.5 and 300 K and in external magnetic fields up to 9 T, a Quantum Design PPMS was used. To probe FE polarization, we measured both, the pyrocurrent at fixed magnetic field and the magnetocurrent at fixed temperature using a Keithley 6517A electrometer. To align the FE domains on cooling, electric poling fields of the order 1 kV/cm were applied.

**Acknowledgement**

This research was partly supported by the Deutsche Forschungsgemeinschaft (DFG) via the Transregional Collaborative Research Center TRR80 (Augsburg, Munich) and the BMBF via ENREKON 03EK3015. We thank Manfred Fiebig for very useful discussions.

**References**

[1] M. Mostovoy, Multiferroic propellers, *Physics* **5**, 16 (2012).
[2] T. Kimura, T. Goto, H. Shintani, K. Ishizaka, T. Arima, and Y. Tokura, Magnetic control of ferroelectric polarization, *Nature* **42**, 55 (2003).
[3] H. Katsura, N. Nagaosa, and A. V. Balatsky, Spin current and magnetoelectric effect in noncollinear magnets, *Phys. Rev. Lett.* **95**, 057205 (2005).
[4] M. Mostovoy, Ferroelectricity in spiral magnets, *Phys. Rev. Lett.* **96**, 067601 (2006).
[5] I. A. Sergienko and E. Dagotto, Role of the Dzyaloshinskii-Moriya interaction in multiferroic perovskites, *Phys. Rev. B* **73**, 094434 (2006).
[6] M. Fiebig, Revival of the magnetoelectric effect, *J. Phys. D* **38**, R123 (2005).

[7] S.-W. Cheong and M. Mostovoy, Multiferroics: a magnetic twist for ferroelectricity, *Nat. Mater.* **6**, 13 (2007).

[8] R. Ramesh and N. A. Spaldin, Multiferroics: progress and prospects in thin films, *Nat. Mater.* **6**, 21 (2007).

[9] O. A. Starykh, Unusual ordered phases of highly frustrated magnets: a review, *Rep. Prog. Phys.* **78**, 052502 (2015).

[10] J. Villain, Two-level systems in a spin-glass model. I. General formalism and two-dimensional model, *J. Phys. C: Sol. Stat. Phys.* **10**, 4793 (1977).

[11] A. V. Chubukov, Chiral, nematic, and dimer states in quantum spin chains, *Phys. Rev. B* **44**, 4693 (1991).

[12] J. Villain in Proceedings of the 13th IUPAP Conference on Statistical Physics, *Ann. Isr. Phys. Soc.* **2**, 565 (1978).

[13] S. Onoda and N. Nagaosa, Chiral spin pairing in helical magnets, *Phys. Rev. Lett.* **99**, 027206 (2007).

[14] T. Hikihara, M. Kaburagi, and Kawamura, Ground-state phase diagrams of frustrated spin-S XXZ chains: Chiral ordered phases, *Phys. Rev. B* **63**, 174430 (2001).

[15] A. Kolezhuk and T. Vekua, Field-induced chiral phase in isotropic frustrated spin chains, *Phys. Rev. B* **72**, 094424 (2005).

[16] F. Heidrich-Meisner, A. Honecker, and T. Vekua, Frustrated ferromagnetic spin-1/2 chain in a magnetic field: The phase diagram and thermodynamic properties, *Phys. Rev. B* **74**, 020403(R) (2006).

[17] I. P. McCulloch, R. Kube, M. Kurz, A. Kleine, U. Schollwöck, and A. K. Kolezhuk, Vector chiral order in frustrated spin chains, *Phys. Rev. B* **77**, 094404 (2008).

[18] T. Hikihara, L. Kecke, T. Momoi, and A. Furusaki, Vector chiral and multipolar orders in the spin-1/2 frustrated ferromagnetic chain in magnetic field, *Phys. Rev. B* **78**, 144404 (2008).

[19] J. Sudan, A. Lüscher, and A. M. Läuchli, Emergent multipolar spin correlations in a fluctuating spiral: The frustrated ferromagnetic spin-1/2 Heisenberg chain in a magnetic field, *Phys. Rev. B* **80**, 140402(R) (2009).

[20] F. Cinti, A. Rettori, M. G. Pini, M. Mariani, E. Micotti, A. Lascialfari, N. Papinutto, A. Amato, A. Caneschi, D. Gatteschi, and M. Affronte, Two-Step Magnetic Ordering in Quasi-One-Dimensional Helimagnets: Possible Experimental Validation of Villain's Conjecture about a Chiral Spin Liquid Phase, *Phys. Rev. Lett.* **100**, 057203 (2008).

[21] L. E. Svistov, T. Fujita, H. Yamaguchi, S. Kimura, K. Omura, A. Prokofiev, A. I. Smirnova, Z. Honda, and M. Hagiwara, New high magnetic field phase of the frustrated S = 1/2 chain compound $LiCuVO_4$, *JETP Letters* **93**, 21 (2011).

[22] M. Mourigal, M. Enderle, B. Fak, R. K. Kremer, J. M. Law, A. Schneidewind, A. Hiess, and A. Prokofiev, Evidence of a Bond-Nematic Phase in $LiCuVO_4$, *Phys. Rev. Lett.* **109**, 027203 (2012).

[23] K. Nawa, M. Takigawa, M. Yoshida, and K. Yoshimura, Anisotropic Spin Fluctuations in the Quasi One-Dimensional Frustrated Magnet $LiCuVO_4$, *J. Phys. Soc. Jpn.* **82**, 094709 (2013).

[24] N. Büttgen, K. Nawa, T. Fujita, M. Hagiwara, P. Kuhns, A. Prokofiev, A. P. Reyes, L. E. Svistov, K. Yoshimura, and M. Takigawa, Search for a spin-nematic phase in the quasi-one-dimensional frustrated magnet $LiCuVO_4$, *Phys. Rev. B* **90**, 134401 (2014).

[25] A. Orlova, E. L. Green, J. M. Law, D. I. Gorbunov, G. Chanda, S. Krämer, M. Horvatić, R. K. Kremer, J. Wosnitza, and G. L. J. A. Rikken, Nuclear Magnetic Resonance Signature of the Spin-Nematic Phase in $LiCuVO_4$ at High Magnetic Fields, *Phys. Rev. Lett.* **118**, 247201 (2017).

[26] M. Enderle, C. Mukherjee, B. Fåk, R.K. Kremer, J.-M. Broto, H. Rosner, S.-L. Drechsler, J. Richter, J. Malek, A. Prokofiev, W. Assmus, S. Pujol, J.-L. Raggazzoni, H. Rakoto, M. Rheinstädter, H.M. Rønnow, Quantum helimagnetism of the frustrated spin-½ chain $LiCuVO_4$, *Europhys. Lett.* **70**, 237 (2005).

[27] S. Nishimoto, S.-L. Drechsler, R. Kuzian, J. Richter, J. Málek, M. Schmitt, J. van den Brink, and H. Rosner, The strength of frustration and quantum fluctuations in $LiVCuO_4$, *Europhys. Lett.* **98**, 37007 (2012).

[28] H.-J. Koo, Ch. Lee, M.-H. Whangbo, G. J. McIntyre, and R. K. Kremer, On the Nature of the Spin Frustration in the $CuO_2$ Ribbon Chains of $LiCuVO_4$: Crystal Structure Determination at 1.6 K, Magnetic Susceptibility Analysis, and Density Functional Evaluation of the Spin Exchange Constants, *Inorg. Chem.* **50**, 3582 (2011).

[29] A. N. Vasil'ev, L. A. Ponomarenko, H. Manaka, I. Yamada, M. Isobe, and Y. Ueda, Magnetic and resonant properties of quasi-one-dimensional antiferromagnet $LiCuVO_4$, *Phys. Rev. B* **64**, 024419 (2001).

[30] H.-A. Krug von Nidda, L. E. Svistov, M. V. Eremin, R. M. Eremina, A. Loidl, V. Kataev, A. Validov, A. Prokofiev, and W. Aßmus, Anisotropic exchange in $LiCuVO_4$ probed by ESR, *Phys. Rev. B* **65**, 134445 (2002).

[31] Y. Naito, K. Sato, Y. Yasui, Y. Kobayashi, and M. Sato, Ferroelectric transition induced by the incommensurate magnetic ordering in $LiCuVO_4$, *J. Phys. Soc. Jpn.* **76**, 023708 (2007).

[32] F. Schrettle, S. Krohns, P. Lunkenheimer, J. Hemberger, N. Büttgen, H.-A. Krug von Nidda, A. V. Prokofiev, and A. Loidl, Switching the ferroelectric polarization by external magnetic fields in the spin = 1/2 chain cuprate $LiCuVO_4$, *Phys. Rev. B* **77**, 144101 (2008).

[33] D. C. Johnston, R. K. Kremer, M. Troyer, X. Wang, A. Klümper, S. L. Bud'ko, A. F. Panchula, and P. C. Canfield, Thermodynamics of spin S=1/2 antiferromagnetic uniform and alternating-exchange Heisenberg chains, *Phys. Rev. B* **61**, 9558 (2000).

[34] M. Yamaguchi, T. Furuta, and M. Ishikawa, Calorimetric study of several cuprates with restricted dimensionality, *J. Phys. Soc. Jpn.* **65**, 2998 (1996).

[35] A. Ruff, S. Krohns, P. Lunkenheimer, A. Prokofiev, and A. Loidl, Dielectric properties and electrical switching behaviour of the spin-driven multiferroic $LiCuVO_4$, *J. Phys.: Condens. Matter* **26**, 485901 (2014).

[36] T. Shang, E. Canévet, M. Morin, D. Sheptyakov, M. T. Fernández-Díaz, E. Pomjakushina, M. Medarde, Design of magnetic spirals in layered perovskites: Extending the stability range far beyond room temperature, *Sci. Adv.* **4**, eaau6386 (2018).

[37] A. Scaramucci, H. Shinaoka, M. V. Mostovoy, M. Müller, C. Mudry, M. Troyer, and N. A. Spaldin, Multiferroic Magnetic Spirals Induced by Random Magnetic Exchanges, *Phys. Rev. X* **8**, 011005 (2018).

[38] A. V. Prokofiev, D. Wichert, and W. Aßmus, Crystal growth of the quasi-one dimensional spin-magnet $LiCuVO_4$, *J. Cryst. Growth* **220**, 345 (2000).

[39] N. Büttgen, H.-A. Krug von Nidda, L. E. Svistov, L. A. Prozorova, A. Prokofiev, and W. Aßmus, Spin-modulated quasi-one-dimensional antiferromagnet $LiCuVO_4$, *Phys. Rev. B* **76**, 014440 (2007).

# Supplementary Information

## Chirality-driven ferroelectricity in LiCuVO$_4$

Alexander Ruff[1], Peter Lunkenheimer[1], Hans-Albrecht Krug von Nidda[1], Sebastian Widmann[1], Andrey Prokofiev[2], Leonid Svistov[3], Alois Loidl[1], Stephan Krohns[1,*]

[1] Experimental Physics V, Center for Electronic Correlations and Magnetism, University of Augsburg, 86135 Augsburg, Germany
[2] Institute of Solid State Physics, Vienna University of Technology, A-1040 Wien, Austria
[3] P.L. Kapitza Institute for Physical Problems, RAS, Moscow 119334, Russia

## 1. Results of heat-capacity

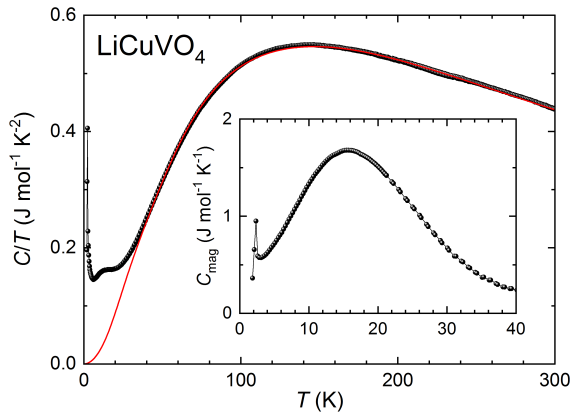

**Figure S1: Heat capacity of LiCuVO$_4$.** Temperature dependence of the specific heat in zero external magnetic fields, plotted as C/T. The red line denotes the phonon contribution with a model assuming one Debye and two Einstein oscillators. The inset shows magnetic contribution $C_{mag}$ to the specific heat for temperatures below 40 K, which has been derived by subtracting the phonon-derived heat capacity

Figure S1 shows the temperature dependence of the specific heat of LiCuVO$_4$ in a representation C/T *vs.* T from 2 K up to room temperature. C/T exhibits a peak at the antiferromagnetic phase transition at 2.3 K followed by a weak local maximum close to 15 K and a broad maximum centred at 140 K. For T > 40 K the specific heat can well be approximated by pure phonon contributions assuming one Debye mode with a Debye temperature of 205 K (weight = 1) and two Einstein oscillators with Einstein temperatures of 340 K (weight = 2.7) and 710 K (weight = 2.9), respectively. Thus, the experimental ratio of the weights of Debye : Einstein oscillators amounts 1 : 5.6 close to the theoretically expected value given by the corresponding ratio of degrees of freedom of acoustical versus optical phonons [3 : 3(n-1) = 1 : 6, where n = 7 denotes the number of atoms in the formula unit of LiCuVO$_4$]. Subtraction of the phonon-derived heat capacity yields the magnetic contribution to the specific heat, as shown in the inset of Fig. S1. Around 15 K one clearly recognizes the characteristic maximum of an antiferromagnetic spin chain in agreement with literature[1]. The temperature of the maximum is determined by the leading exchange interaction. In case of two competing exchange interactions within the chains the maximum in C(T) shifts to lower temperature[2]. It is important to note that the vector chiral (VC) phase and the long-range ordered ferroelectric (FE) phase show up well below this specific-heat maximum. This indicates that only in case of one-dimensional spin chains with competing exchange interactions chiral phases can form giving rise to polar order.

## 2. Results of pyrocurrent measurements

Figure S2 shows the temperature dependent polarisation as determined from pyrocurrent experiments. The measurements were performed along the **a**-direction for applied magnetic fields along the **b**- (a) and along the **c**-direction (b). In zero magnetic fields a strong increase in polarisation at $T_N$ signals the onset of spin-driven ferroelectricity. The measured saturation polarisation in the order of 20 - 30 µC/m$^2$ agrees well with previously reported values[3,4] for the FE order due to the formation of a 3D long-range ordered cycloidal spin spiral, which propagates along **b**. As described in detail in ref. 4, the polarisation for applied magnetic fields along **b** vanishes exceeding the first critical magnetic field ($H_{c1} \approx 3T$). In this case, the chiral vector **e** is along the propagation of the spin-spiral (i.e., **Q**) denoting the helical state, for which, following **P** $\propto$ **e** $\times$ **Q**, the ferroelectric polarisation **P** vanishes. A small increase at $T_N$ for 4 T and 5 T indicates a slight misalignment of the sample in the magnetic field. However, no signature of the VC phase at 11 K is detected in applied magnetic fields up to 9 T (for H||b). Figure S2 (b) documents the polarisation for H||c, which resembles the measurement of ref. 4 at low temperatures (T < 5 K). At 11 K polarisation evolves as discussed in the main paper, which increases with increasing magnetic fields. The saturation polarisation of the VC phase is always well below the polarisation induced by the spin spiral phase. As discussed in the main text, this results from the fact that in the VC phase the twist angles of the given chirality vary statistically, while in the spin-spiral phase all twist angles are strictly correlated. The inset depicts the impact of an applied electric field on the pyrocurrent while cooling (an applied magnetic field of 7 T along **c**-direction). Only, in the case of electric-field poling a macroscopic polarisation is induced in the VC as well as in the spin-spiral phase.

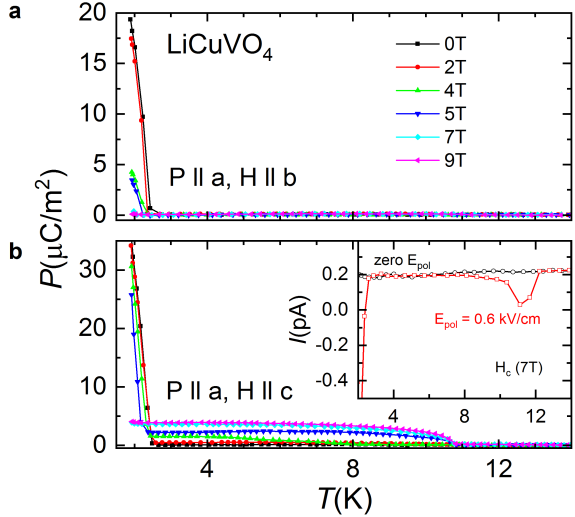

**Figure S2: Pyrocurrent measurements of LiCuVO$_4$.** (a) Temperature dependence of the electric polarization along the crystallographic **a** direction, **P**||**a**, for a series of magnetic fields $\mu_0 H$, between 0 and 9 T with **H**||**b**. (b) Temperature dependence of the electric polarization along the crystallographic **a** direction, **P**||**a**, for a series of magnetic fields $\mu_0 H$, between 0 and 9 T with **H**||**c**. The inset shows the measured pyrocurrent while heating the sample. The black symbols denote measurements without electric-field cooling, while the red symbols indicate the measurement for applied electric poling fields $E_{pol}$ = 0.6 kV/cm.

**References**

[1] L. A. Prozorova, S. S. Sosin, L. E. Svistov, N. Büttgen, J. B. Kemper, A. P. Reyes, S. Riggs, A. Prokofiev, O. A. Petrenko, Magnetic field driven 2D-3D crossover in the S = ½ frustrated chain magnet LiCuVO$_4$, *Phys. Rev. B* **91**, 174410 (2015).

[2] J. Sirker, Thermodynamics of multiferroic spin chains, *Phys. Rev. B* **81**, 014419 (2010).

[3] Y. Yasui, Y. Naito, K. Sato, T. Moyoshi, M. Sato, and K. Kakurai, Relationship between Magnetic Structure and Ferroelectricity of LiVCuO$_4$, *J. Phys. Soc. Jpn.* **77**, 023712 (2008).

[4] F. Schrettle, S. Krohns, P. Lunkenheimer, J. Hemberger, N. Büttgen, H.-A. Krug von Nidda, A. V. Prokofiev, and A. Loidl, Switching the ferroelectric polarization by external magnetic fields in the spin = 1/2 chain cuprate LiCuVO$_4$, *Phys. Rev. B* **77**, 144101 (2008).